\documentclass[prd,aps,showpacs,nofootinbib,preprint]{revtex4}
\usepackage{graphicx,color,amsmath,amsxtra}
\usepackage{epsf}
\usepackage{amssymb}
\usepackage{enumerate}
\usepackage{hhline}
\usepackage{array}
\usepackage{tabularx}

\newcommand{\bear}{\begin{array}}  \newcommand{\eear}{\end{array}}
\newcommand{\bea}{\begin{eqnarray}}  \newcommand{\eea}{\end{eqnarray}}
\newcommand{\beq}{\begin{equation}}  \newcommand{\eeq}{\end{equation}}
\newcommand{\bef}{\begin{figure}}  \newcommand{\eef}{\end{figure}}
\newcommand{\bec}{\begin{center}}  \newcommand{\eec}{\end{center}}

\newcommand{\Eqn}[1]{&\hspace{-0.2em}#1\hspace{-0.2em}&}
\def\Vec#1{\mbox{\boldmath $#1$}}

\def\be{\begin{equation}}
\def\ee{\end{equation}}
\def\bea{\begin{eqnarray}}
\def\eea{\end{eqnarray}}
\def\beq{\begin{eqnarray}}
\def\eeq{\end{eqnarray}}

\def\be{\begin{equation}}
\def\ee{\end{equation}}
\def\bea{\begin{eqnarray}}
\def\eea{\end{eqnarray}}
\def\beq{\begin{eqnarray}}
\def\eeq{\end{eqnarray}}

\begin{document}

\title{Oscillating phantom in $F(R)$ gravity
}

\author{Kazuharu Bamba\footnote{E-mail address: bamba@phys.nthu.edu.tw} 
and Chao-Qiang Geng\footnote{E-mail address: geng@phys.nthu.edu.tw}
}
\affiliation{
Department of Physics, National Tsing Hua University, Hsinchu, Taiwan 300
}

%\date{\today}

%%%%%%%%%%%%%%%%%%%%%
%  Abstract
%%%%%%%%%%%%%%%%%%%%%
\begin{abstract} 
We investigate the oscillating effective equation of state (EoS) of the 
universe around the phantom divide in the framework of $F(R)$ gravity. 
We illustrate the behavior of $F(R)$ with realizing 
multiple crossings of the phantom divide. 
\end{abstract}
%%%%%%%%%%%%%%%%%%%%%

%----------------------------
%\preprint{}
%\hspace{13.0cm}
%----------------------------

\maketitle
%==============================================================================

%%%%%%%%%%%%%%%%%%%%%%%%%%%
%%%  (Sec. I)
%%%%%%%%%%%%%%%%%%%%%%%%%%%
%\section{Introduction}

Recent observations have supported that the current expansion of the universe 
is accelerating~\cite{WMAP, SN1}. 
There are two broad categories to account for this scenario 
\cite{Peebles, Copeland:2006wr, DM, Nojiri:2006ri, rv-2}. 
One is the introduction of ``dark energy'' in the framework of general 
relativity. The other is the study of a modified gravitational theory, e.g., 
$F(R)$ gravity, in which the action is represented by an 
arbitrary function $F(R)$ of the scalar curvature $R$ (for reviews, 
see~\cite{Nojiri:2006ri, rv-2}). 
It is also known that,
according to the observational 
data~\cite{observational status}, the ratio of the effective pressure to the 
effective energy density of the universe, 
i.e., the effective equation of state (EoS) 
$w_\mathrm{eff}\equiv p_\mathrm{eff}/\rho_\mathrm{eff}$, 
may evolve from larger than $-1$ (non-phantom phase) to less 
than $-1$ (phantom phase~\cite{phantom}). This means that it crosses $-1$ 
(the phantom divide) at the present time or in near future. 
There are many models to realize the crossing of the phantom divide 
(for a detailed review, see~\cite{Copeland:2006wr}). 
In the framework of $F(R)$ 
gravity~\cite{Nojiri:2006ri, Abdalla:2004sw, Amendola:2007nt, 
Brevik}, an explicit model with realizing a crossing 
of the phantom divide has been constructed in Ref.~\cite{Bamba:2008hq} and 
its thermodynamics  has been 
examined~\cite{Bamba:2009ay}. 
In this model, only one crossing of the phantom divide can occur because 
there appears the Big Rip singularity~\cite{BR} 
at the end of the phantom phase. 
However, in the framework of general 
relativity multiple crossings of the phantom divide can be realized, e.g., in 
an oscillating quintom model~\cite{Feng:2004ff} or a quintom with two 
scalar fields~\cite{Zhang:2005eg}. 

In this paper, we study the oscillating effective EoS of the universe around 
the phantom divide in the framework of $F(R)$ gravity. 
We investigate the behavior of $F(R)$ with realizing multiple 
crossings of the phantom divide. 
We use units of $k_\mathrm{B} = c = \hbar = 1$ and denote the
gravitational constant $8 \pi G$ by 
${\kappa}^2 \equiv 8\pi/{M_{\mathrm{Pl}}}^2$ 
with the Planck mass of $M_{\mathrm{Pl}} = G^{-1/2} = 1.2 \times 10^{19}$GeV.

%%%%%%%%%%%%%%%%%%%%%%%%%%%
%%%  (Sec. II)
%%%%%%%%%%%%%%%%%%%%%%%%%%%
%\section{Reconstruction method}

We start the reconstruction method of $F(R)$ gravity proposed in Ref.~\cite{RM}. 
The action of $F(R)$ gravity with matter is as follows: 
\begin{eqnarray}
S = \int d^4 x \sqrt{-g} \left[ \frac{F(R)}{2\kappa^2} +
{\mathcal{L}}_{\mathrm{matter}} \right]\,,
\label{eq:1}
\end{eqnarray}
where $g$ is the determinant of the metric tensor $g_{\mu\nu}$ and
${\mathcal{L}}_{\mathrm{matter}}$ is the matter Lagrangian. 
By using proper functions $P(\phi)$ and $Q(\phi)$ of a scalar field $\phi$, 
the action in Eq.~(\ref{eq:1}) can be rewritten to 
\begin{eqnarray}
S=\int d^4 x \sqrt{-g} \left\{ \frac{1}{2\kappa^2} \left[ P(\phi) R + Q(\phi)
\right] + {\mathcal{L}}_{\mathrm{matter}} \right\}\,.
\label{eq:2}
\end{eqnarray}
The scalar field $\phi$ may be regarded as an auxiliary scalar field because 
it has no kinetic term. From Eq.~(\ref{eq:1}), the equation of 
motion of $\phi$ is given by 
\begin{eqnarray}
0=\frac{d P(\phi)}{d \phi} R + \frac{d Q(\phi)}{d \phi}\,.
\label{eq:3}
\end{eqnarray}
Substituting $\phi=\phi(R)$ into the action in Eq.~(\ref{eq:2}) yields 
the expression of $F(R)$ as 
\begin{eqnarray}
F(R) = P(\phi(R)) R + Q(\phi(R))\,.
\label{eq:4}
\end{eqnarray}
 From Eq.~(\ref{eq:2}), the field equation of modified gravity is 
derived as
\begin{eqnarray}
\frac{1}{2}g_{\mu \nu} \left[ P(\phi) R + Q(\phi) \right]
-R_{\mu \nu} P(\phi) -g_{\mu \nu} \Box P(\phi) +
{\nabla}_{\mu} {\nabla}_{\nu}P(\phi) + \kappa^2
T^{(\mathrm{matter})}_{\mu \nu} = 0\,,
\label{eq:5}
\end{eqnarray}
where ${\nabla}_{\mu}$ is the covariant derivative operator associated with 
$g_{\mu \nu}$, $\Box \equiv g^{\mu \nu} {\nabla}_{\mu} {\nabla}_{\nu}$ 
is the covariant d'Alembertian for a scalar field, and 
$T^{(\mathrm{matter})}_{\mu \nu}$ is the contribution to 
the energy-momentum tensor form matter.

We assume the flat 
Friedmann-Robertson-Walker (FRW) space-time with the metric,
\begin{eqnarray}
{ds}^2 = -{dt}^2 + a^2(t)d{\Vec{x}}^2\,,
\label{eq:6}
\end{eqnarray}
where $a(t)$ is the scale factor. 
In this background, the components of
$(\mu,\nu)=(0,0)$ and $(\mu,\nu)=(i,j)$ $(i,j=1,\cdots,3)$ 
in Eq.~(\ref{eq:5}) read 
\begin{eqnarray}
&&
-6H^2P(\phi(t)) -Q(\phi(t)) -6H \frac{dP(\phi(t))}{dt} + 2\kappa^2\rho = 0\,,
\label{eq:7} \\
&&
2\frac{d^2P(\phi(t))}{dt^2}+4H\frac{dP(\phi(t))}{dt}+
\left(4\dot{H}+6H^2 \right)P(\phi(t)) +Q(\phi(t)) + 2\kappa^2 p = 0\,,
\label{eq:8}
\end{eqnarray} 
where $H=\dot{a}/a$ is the Hubble parameter with 
$\dot{~}=\partial/\partial t$ and 
$\rho$ and $p$ are the sum of the energy density and
pressure of matters with a constant EoS parameter $w_i$, respectively, 
with $i$ being some component of matters. 
After eliminating $Q(\phi)$ from Eqs.~(\ref{eq:7}) and (\ref{eq:8}), we 
obtain
\begin{eqnarray}
\frac{d^2P(\phi(t))}{dt^2} -H\frac{dP(\phi(t))}{dt} +2\dot{H}P(\phi(t)) +
\kappa^2 \left( \rho + p \right) = 0\,.
\label{eq:9}
\end{eqnarray}
The scalar field $\phi$ may be taken as $\phi = t$ 
if it is redefined properly. 
By representing $a(t)$ as
\begin{eqnarray}
a(t) = \bar{a} \exp \left( \tilde{g}(t) \right)
\label{eq:10}
\end{eqnarray}
in terms of a constant of $\bar{a}$ and  a proper function of $\tilde{g}(t)$
and using $H= d \tilde{g}(\phi)/\left(d \phi \right)$, we rewrite 
Eq.~(\ref{eq:9}) to be 
\begin{eqnarray}
&&
\frac{d^2P(\phi)}{d\phi^2} -\frac{d \tilde{g}(\phi)}{d\phi}
\frac{dP(\phi)}{d\phi} +2 \frac{d^2 \tilde{g}(\phi)}{d \phi^2}
P(\phi) \nonumber \\
&& \hspace{10mm}
{}+
\kappa^2 \sum_i \left( 1+w_i \right) \bar{\rho}_i
\bar{a}^{-3\left( 1+w_i \right)} \exp
\left[ -3\left( 1+w_i \right) \tilde{g}(\phi) \right] = 0\,,
\label{eq:11}
\end{eqnarray}
where $\bar{\rho}_i$ is a constant. 
Moreover, from Eq.~(\ref{eq:7}), we get
\begin{eqnarray}
Q(\phi) \Eqn{=} -6 \left[ \frac{d \tilde{g}(\phi)}{d\phi} \right]^2 P(\phi)
-6\frac{d \tilde{g}(\phi)}{d\phi} \frac{dP(\phi)}{d\phi} \nonumber \\
&& \hspace{10mm}
{}+
2\kappa^2 \sum_i \bar{\rho}_i \bar{a}^{-3\left( 1+w_i \right)}
\exp
\left[ -3\left( 1+w_i \right) \tilde{g}(\phi) \right]\,.
\label{eq:12}
\end{eqnarray}

%%%%%
We note that if we redefine the auxiliary scalar field $\phi$
by $\phi=\Phi(\varphi)$ with a proper function $\Phi$ and define 
$\tilde P(\varphi)\equiv P(\Phi(\varphi))$ and
$\tilde Q(\varphi)\equiv Q(\Phi(\varphi))$, the new action 
\begin{eqnarray}
S \Eqn{=} \int d^4 x \sqrt{-g}\left[ \frac{\tilde F(R)}{2\kappa^2} 
+{\cal L}_{\rm matter} \right]\,, 
\label{eq:A1} \\
\tilde F(R) \Eqn{\equiv} \tilde P(\varphi) R + \tilde Q(\varphi)\,, 
\label{eq:A2}
\end{eqnarray}
is equivalent to the action in Eq.~(\ref{eq:2}) because $\tilde F(R) = F(R)$. 
Here, $\varphi$ is the inverse function of $\Phi$ and 
we can solve $\varphi$ with respect to $R$ as 
$\varphi=\varphi(R) = \Phi^{-1}(\phi(R))$ by using $\phi = \phi(R)$. 
As a consequence, we have the choices in $\phi$ like a gauge symmetry 
and thus we can identify $\phi$ with time $t$, i.e., $\phi=t$, which can be 
interpreted as a gauge condition corresponding to the reparameterization of 
$\phi=\phi(\varphi)$~\cite{Bamba:2008hq}. 
Hence, if we have the relation $t = t(R)$, in principle we can obtain the 
form of $F(R)$ by solving Eq.~(\ref{eq:11}) with Eqs.~(\ref{eq:4}) and 
(\ref{eq:12}). 
%%%%%

%%%%%
We also mention that the phantom crossing cannot be described by a naive model 
of $F(R)$ gravity. 
To demonstrate the crossing, $F(R)$ needs to be a double-valued function, 
where the cut could correspond to $w_\mathrm{eff} = -1$. 
However, the crossing can be performed by the extension of $F(R)$ gravity, 
whose action is given by $P(\phi) R + Q(\phi)$. 
%%%%%

%%%%%%%%%%%%%%%%%%%%%%%%%%%
%%%  (Sec. III)
%%%%%%%%%%%%%%%%%%%%%%%%%%%
%\section{Model}

%%%%%
We consider a cosmology at late times. 
%%%%%
To illustrate the behavior of $F(R)$ with realizing multiple 
crossings of the phantom divide, we investigate the case in which the Hubble 
rate $H(t)$ is expressed by an oscillating function, given 
by~\cite{OF} 
\begin{eqnarray}
H \Eqn{=} H_0 +H_1 \sin \nu t\,,
\label{eq:13} \\
\nu \Eqn{=} \frac{3\pi}{2} H_{\mathrm{p}}\,,
\label{eq:14}
\end{eqnarray}
where $H_{\mathrm{p}}=2.13h \times 10^{-42}$GeV~\cite{Kolb and Turner}
with $h =0.70$~\cite{Freedman:2000cf} is the present Hubble parameter. 
Here, we take $H_0 = \alpha H_1$  and $H_1$ to be 
constants with $\alpha >1$ so that $H$ can always be positive and the universe 
can expand. By using Eqs.~(\ref{eq:13}) and  (\ref{eq:14}) and 
$t_p \approx 1/H_{\mathrm{p}}$, the present time $t_{\mathrm{p}}$ is expressed 
by $t_{\mathrm{p}} = \left(3\pi/2\right) \nu^{-1}$ and 
$H_1 = H_{\mathrm{p}}/\left(\alpha-1\right)$. 
%%%%%
We note that the behavior of $H$ in Eq.~(\ref{eq:13}) is only acceptable 
at late times. 
%%%%%

In the FRW background, the effective 
energy density and pressure of the universe are given by 
$\rho_\mathrm{eff} = 3H^2/\kappa^2$ and
$p_\mathrm{eff} = -\left(2\dot{H} + 3H^2 \right)/\kappa^2$, respectively. 
The effective EoS $w_\mathrm{eff} = p_\mathrm{eff}/\rho_\mathrm{eff}$ 
is defined as~\cite{Nojiri:2006ri}
\begin{eqnarray}
w_\mathrm{eff} \equiv -1 -\frac{2\dot{H}}{3H^2}\,, 
\label{eq:15}
\end{eqnarray}
which implies that a crossing of the phantom divide 
occurs when the sign of $\dot{H}$ changes. 
In case of Eq.~(\ref{eq:13}), multiple crossings of the phantom divide can 
be realized. This means that the phantom phase is transient like the 
model in Ref.~\cite{Abdalla:2004sw} and hence there is no Big Rip singularity. 
Hereafter, we take $\phi = t$. 
We show the time evolution of $w_\mathrm{eff}$ in Fig.~1 with 
$\tilde{t} \equiv \nu t$. 
In all figures, we take $\alpha=10$. 
From Fig.~1, we see that at the present time 
$t_{\mathrm{p}} = \left(3\pi/2\right) \nu^{-1}$, 
the universe enters the phantom phase $w_\mathrm{eff} < -1$ from the 
non-phantom phase $w_\mathrm{eff} > -1$. Thus, a crossing of the phantom 
divide occurs.

%%%%%% Fig. 1 %%%%%%%%%
\begin{figure}[tbp]
\begin{center}
%\resizebox{!}{8cm}{
   \includegraphics{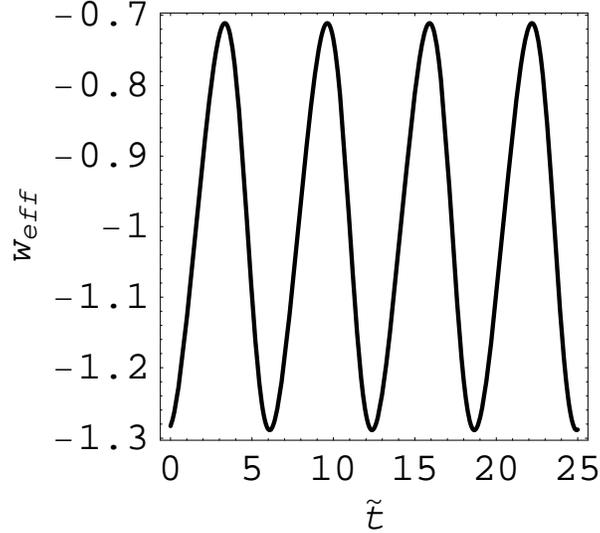}
%                  }
\caption{Time evolution of $w_\mathrm{eff}$ for $\alpha = 10$ 
with $\tilde{t} = \nu t$. 
}
\end{center}
\label{fg:1}
\end{figure}
%%%%%%%%%%%%%%%%%%%%%%%%

 From Eqs.~(\ref{eq:10}) and (\ref{eq:13}) with $H_0 = \alpha H_1$, 
$H= d \tilde{g}(t)/\left(d t \right)$ and $R=6\left( \dot{H} + 2H^2 \right)$, 
we obtain 
\begin{eqnarray}
\tilde{g}(t) \Eqn{=} H_1 \left( \alpha t -\frac{1}{\nu} \cos \nu t \right)\,, 
\label{eq:16} \\
a(t) \Eqn{=} \exp \left[ -\frac{\alpha}{\alpha -1} + 
H_1 \left( \alpha t -\frac{1}{\nu} \cos \nu t \right) \right]\,,
\label{eq:17} 
\end{eqnarray}
\begin{eqnarray}
%\\
R 
%\Eqn{=}
= 6 H_1^2 \left[ \frac{\nu}{H_1} \cos \nu t + 
2 \left( \alpha + \sin \nu t \right)^2 \right]\,,
\label{eq:18}
\end{eqnarray}
where we have taken $\bar{a}= \exp \left[ -\alpha\left(\alpha -1\right) 
\right]$ so that the present value of the scale factor 
should be unity.

We define $X \equiv \cos \nu t$ and solve Eq.~(\ref{eq:18}) with 
respect to $X$. If $\alpha$ is much larger than unity, we can neglect
the term proportional to $\sin^2 \nu t$ in Eq.~(\ref{eq:18}) and therefore 
obtain the approximate solutions 
\begin{eqnarray}
X (\tilde{R}) \approx \frac{\beta}{16 \alpha^2 \beta^2 +1} 
\left[ -2\alpha^2 + \frac{\tilde{R}}{6\beta^2} \pm 4\alpha 
\sqrt{18 \alpha^2 \beta^2 +1-\frac{\tilde{R}}{6}}
\right]\,,
\label{eq:19}
\end{eqnarray}
where
\begin{eqnarray}
\tilde{R} \Eqn{=} \frac{R}{\nu^2}\,, \quad 
\beta = \frac{H_1}{\nu}\,.
\label{eq:20}
\end{eqnarray} 
In what follows, we use the upper sign in Eq.~(\ref{eq:19}). 

For simplicity, we consider the case in which 
there exists a matter with a constant EoS parameter $w=p/\rho$. 
In this case, 
Eqs.~(\ref{eq:11}) and (\ref{eq:12}) are rewritten to 
\begin{eqnarray}
&&
18 \beta \left( \sqrt{1-X^2} \mp 4 \alpha \beta X \right)^2 
\frac{d^2 P(\tilde{R})}{d \tilde{R}^2}
\nonumber \\
&& 
{}-3\left[ \left( 4 \alpha^2 \beta^2 +1 \right)X \pm \alpha \beta 
\left( 4\beta X +3 \right) \sqrt{1-X^2} -\beta \left( 1-X^2 \right) 
\right] \frac{d P(\tilde{R})}{d \tilde{R}} +X P(\tilde{R}) 
\nonumber \\
&& 
{}+\frac{\kappa^2 }{2H_1 \nu}
(1+w) \bar{\rho} \exp \left\{ -3\left( 1+w \right) 
\left[ 
-\frac{\alpha}{\alpha -1} +\beta \left( \alpha \arccos X - X 
\right)
\right]
\right\} =0\,
\label{eq:21}
\end{eqnarray}
and
\begin{eqnarray}
\frac{Q(\tilde{R})}{\nu^2} \Eqn{=} 
-6\beta^2 \left( \alpha \pm \sqrt{1-X^2} \right)^2 P(\tilde{R}) 
-36\beta^2 \left( \alpha \pm \sqrt{1-X^2} \right) 
\left( 4\alpha \beta X \mp \sqrt{1-X^2} \right) 
\frac{d P(\tilde{R})}{d \tilde{R}} 
\nonumber \\
&&
{}+ \frac{2\kappa^2}{\nu^2}
\bar{\rho} \exp \left\{ -3\left( 1+w \right) 
\left[ -\frac{\alpha}{\alpha -1} +\beta \left( \alpha \arccos X - X 
\right) \right] \right\}\,,
\label{eq:22}
\end{eqnarray}
respectively, where the upper signs correspond to 
$\sin \nu t = +\sqrt{1-X^2}$, while  
the lower ones  $\sin \nu t = -\sqrt{1-X^2}$. 
We will concentrate on the latter case, 
while the former one will be remarked at the end. 
Here, $\bar{\rho}$ corresponds to the 
present energy density of the matter. In particular, 
we use the present value of the 
cold dark matter with $w=0$ for $\bar{\rho}$, i.e., 
$\bar{\rho} = 0.233 \rho_\mathrm{c}$~\cite{WMAP}, 
where $\rho_\mathrm{c} =3H_0^2/\left(8 \pi G \right) =
3.97 \times 10^{-47} \mathrm{GeV}^4$ is the critical energy density. From 
Eq.~(\ref{eq:4}) and the first relation in Eq.~(\ref{eq:20}), we have 
\begin{eqnarray}
\frac{F(\tilde{R})}{2\kappa^2} = \frac{\nu^2}{2\kappa^2}
\left( P(\tilde{R}) \tilde{R} + \frac{Q(\tilde{R})}{\nu^2} \right)\,.
\label{eq:23}
\end{eqnarray}
To examine $F(\tilde{R})$, we numerically solve 
Eqs.~(\ref{eq:21})--(\ref{eq:23}).

%%%%%%%%%%%%%%%%%%%%%%%
%%%  Figures
%%%%%%%%%%%%%%%%%%%%%%%
%%%%%% Fig. 2 %%%%%%%%%
\begin{figure}[tbp]
\begin{center}
%\resizebox{!}{8cm}{
   \includegraphics{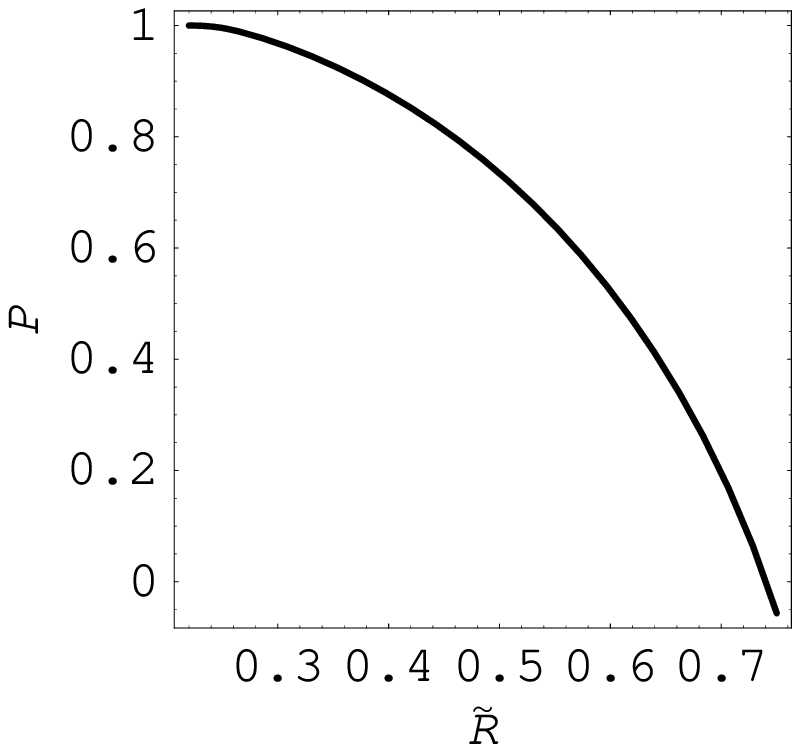}
   \includegraphics{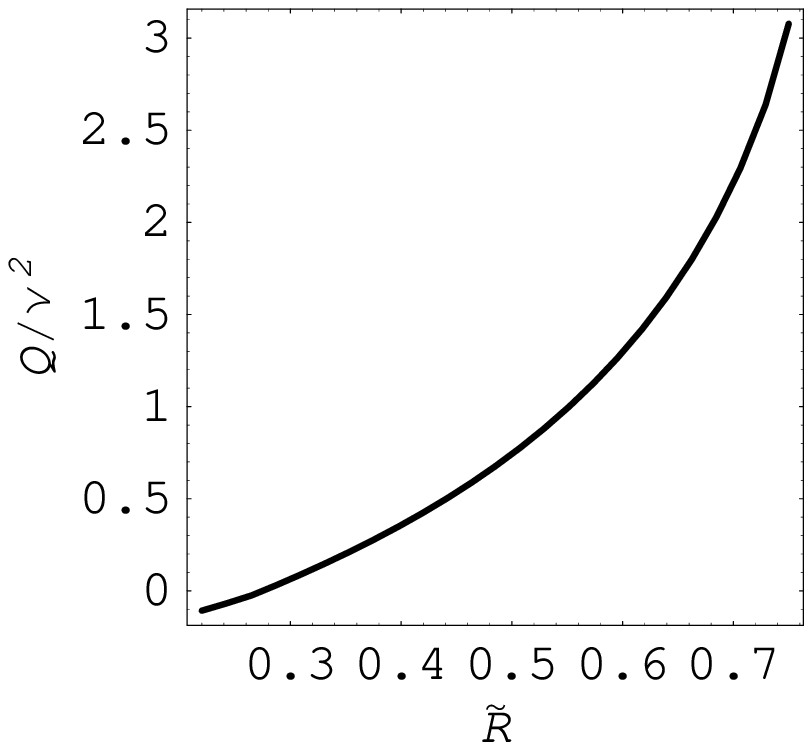}
%                  }
\caption{
$P(\tilde{R})$ and $Q(\tilde{R})/\nu^2$ as  functions of $\tilde{R}$ for $\alpha = 10$ and 
$\bar{\rho} = 0.233 \rho_\mathrm{c}$. 
}
\end{center}
\label{fg:2}
\end{figure}
%%%%%%%%%%%%%%%%%%%%%%

In Fig.~2, we depict $P(\tilde{R})$ and $Q(\tilde{R})/\nu^2$ as 
functions of $\tilde{R}$. 
The range of $\tilde{R}$ is given by $0.22 \leq \tilde{R} \leq 0.75$
by solving Eq.~(\ref{eq:21}) numerically, corresponding to $-1 < X < 1$ 
in Eq.~(\ref{eq:19}). 
Here, we have taken the initial conditions as $P(\tilde{R} = 0.22) =1.0$ and 
$d P(\tilde{R} = 0.22)/ ( d \tilde{R} ) =0$ so that at a smaller curvature,
$F(R)/\left( 2\kappa^2 \right)$ could contain the term 
$R/\left( 2\kappa^2 \right)$, i.e., the ordinary Einstein-Hilbert action. 
By using Eq.~(\ref{eq:23}), 
we show the behavior of $F(\tilde{R})/\left(2 \kappa^2 \right)$ in Fig.~3. 
%%%%%
We note that the qualitative behavior of 
$F(\tilde{R})/\left(2 \kappa^2 \right)$ in Fig.~3 
does not depend on the initial conditions and that 
the quantitative values of $F(\tilde{R})/\left(2 \kappa^2 \right)$ do not also 
depend on the initial conditions strongly. 
%%%%%
Furthermore, we illustrate $w_\mathrm{eff} (\tilde{R}) =
-1 -\left[2/\left(3\beta \right) \right] X(\tilde{R})/\left( \alpha 
-\sqrt{1-X^2(\tilde{R})} \right)^2$ in Fig.~4. 
The time evolution of $\tilde{R}$ is given in Fig.~5. 
 From Figs.~4 and 5, we see that multiple crossings of the phantom divide can 
be realized. 
We note that the results in the all figures are shown by dimensionless 
quantities. 
We also note that the lower sign in Eq.~(\ref{eq:19}) leads to 
$0.59 \leq \tilde{R} \leq 1.11$ with the results being qualitatively 
the same as those obtained by the upper sign above. 

From Fig.~3, we see that $F(\tilde{R})$ increases in terms of $\tilde{R}$ 
around the present curvature 
$\tilde{R} (t=t_{\mathrm{p}}) = 12\left[2/\left( 3\pi \right)\right]^2 = 
0.54$. This behavior is reasonable because 
in the Hu-Sawicki model~\cite{Hu:2007nk} of $F(R)$ gravity, which 
passes the solar system tests, $F(R)$ increases around the present 
curvature. 
%%
%%Thus, $F(\tilde{R})$ in Fig.~3 is consistent with the Hu-Sawicki model. 
%%
We mention that such a behavior is typical for a general class of viable 
$F(R)$ gravities introduced in Ref.~\cite{Cognola:2007zu} to which 
the Hu-Sawicki model belongs. This class of modified gravities can satisfy 
the solar system tests and unify inflation with the late-time cosmic 
acceleration. 
%%%%%
As viable models of $F(R)$ gravity, e.g., the models in 
Refs.~\cite{Starobinsky:2007hu, Appleby:2007vb, 
Amendola:2006we, Li:2007xn, Tsujikawa:2007xu, Capozziello:2007eu} are also 
known (see also~\cite{Miranda:2009rs}). 
%%%%%
%%%%%
We remark that it is impossible to state anything on whether the reconstructed 
$F(R)$ gravity can pass the solar system tests and cosmological 
constraints at early times because the reconstructed $F(R)$ gravity cannot 
describe the behavior of $F(R)$ gravity for large values of $R$. 
%%%%%

%%%%%
We note the stability for the obtained solutions of the phantom crossing 
under a quantum correction coming from the conformal anomaly.  
In Ref.~\cite{Bamba:2008hq}, it has been shown that 
the quantum correction of massless conformally-invariant fields could be small 
when the phantom crossing occurs and therefore the solutions of the phantom 
crossing could be stable under the quantum correction, although the quantum 
correction becomes important near the Big Rip singularity. 
In the present model of $F(R)$ gravity with realizing multiple crossings of 
the phantom divide, the obtained solution can be stable 
because the phantom phases are transient and there is no Big Rip singularity. 
%%%%%

Finally, we would like to remark that for the other possible solution of 
$\sin \nu t = + \sqrt{1-X^2}$, i.e., the case of the upper signs in 
Eqs.~(\ref{eq:21}) and (\ref{eq:22}), $F(\tilde{R})$ approaches 
minus infinity as the curvature becomes larger than the present one although 
it increases in terms of $\tilde{R}$ at smaller curvature than the 
present one. Such a behavior is incompatible with the Hu-Sawicki model, in 
which $F(R) \sim R + const.$ at a much larger curvature than the 
present one.

%%%%%%%%%%%%%%%%%%%%%%%
%%%  Figures
%%%%%%%%%%%%%%%%%%%%%%%
%%%%%% Fig. 3 %%%%%%%%%
\begin{figure}[tbp]
\begin{center}
%\resizebox{!}{8cm}{
   \includegraphics{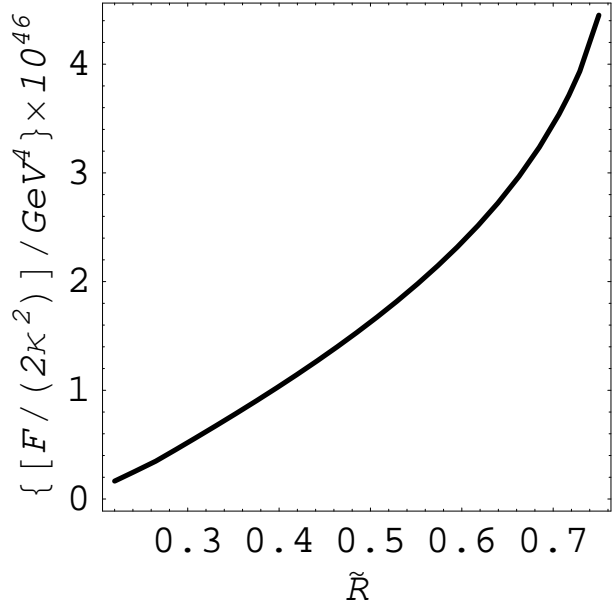}
%                  }
\caption{
Behavior of $F(\tilde{R})/\left(2 \kappa^2 \right)$ as a function of 
$\tilde{R}$. 
Legend is the same as Fig.~2.
}
\end{center}
\label{fg:3}
\end{figure}
%%%%%%%%%%%%%%%%%%%%%%

%%%%%% Fig. 4 %%%%%%%%%
\begin{figure}[tbp]
\begin{center}
%\resizebox{!}{8cm}{
   \includegraphics{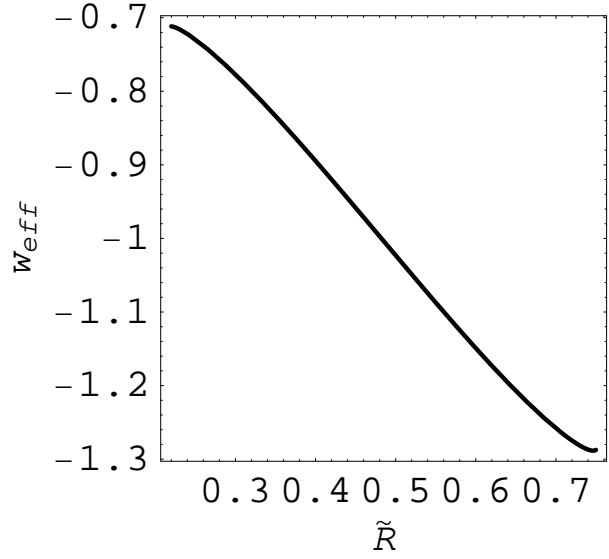}
%                  }
\caption{
Behavior of $w_\mathrm{eff}(\tilde{R})$. 
Legend is the same as Fig.~2.
}
\end{center}
\label{fg:4}
\end{figure}
%%%%%%%%%%%%%%%%%%%%%%

%%%%%% Fig. 5 %%%%%%%%%
\begin{figure}[tbp]
\begin{center}
%\resizebox{!}{8cm}{
   \includegraphics{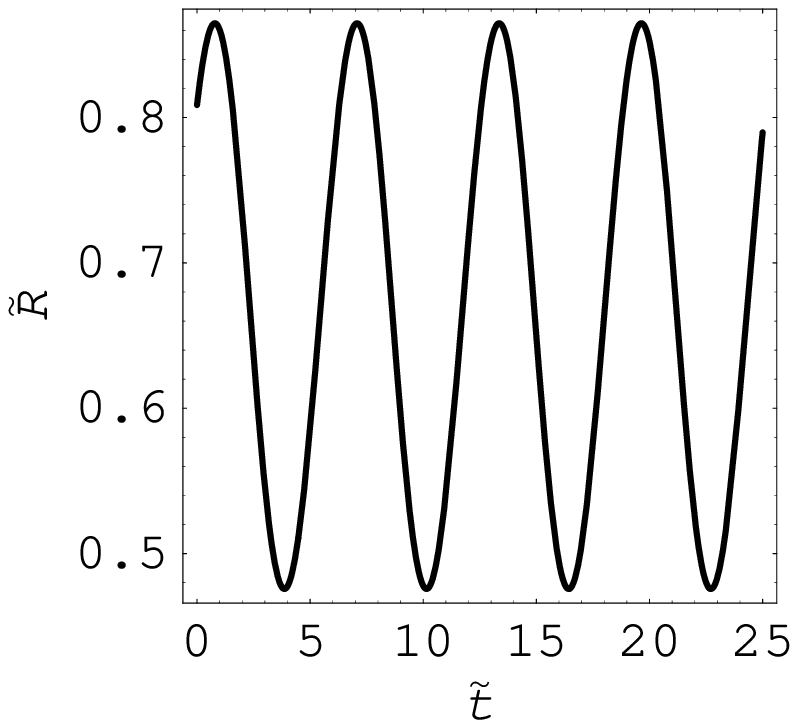}
%                  }
\caption{
Time evolution of $\tilde{R}$. 
Legend is the same as Fig.~1.
}
\end{center}
\label{fg:5}
\end{figure}
%%%%%%%%%%%%%%%%%%%%%%

%%%%%%%%%%%%%%%%%%%
%%%  (Sec. IV)
%%%%%%%%%%%%%%%%%%%
%\section{Conclusion}

In summary, we have studied the oscillating effective EoS of the 
universe around the phantom divide in $F(R)$ gravity. In particular, we 
have analyzed the behavior of $F(R)$ gravity with realizing multiple crossings 
of the phantom divide. 
Our result can be interpreted as an explicit example to 
illustrate that multiple phantom crossings can occur in $F(R)$ gravity as 
the scalar field theories such as an oscillating quintom 
model~\cite{Feng:2004ff} in the framework of general relativity.

%%%%%%%%%%%%%%%%%%%%%%%%
%%%  Acknowledgments
%%%%%%%%%%%%%%%%%%%%%%%%
%\section*{Acknowledgments}

We thank Professor Sergei D. Odintsov and Professor Shin'ichi Nojiri for their 
collaboration in our previous work~\cite{Bamba:2008hq} and important comments. 
We are also grateful to Dr. Tsutomu Kobayashi for useful communications. 
%%%%%
K.B. acknowledges the KEK theory exchange program 
for physicists in Taiwan and the very kind hospitality of 
KEK. 
%%%%%
This work is supported in part by
the National Science Council of R.O.C. under:
Grant \#s: NSC-95-2112-M-007-059-MY3 and
National Tsing Hua University under Grant \#:
97N2309F1 (NTHU).

%%%%%%%%%%%%%%%%%%%%%%%%%%%%%%%%%
%% thebibliography environment
%%%%%%%%%%%%%%%%%%%%%%%%%%%%%%%%%

\end{document}